\documentclass[preprint,5p]{elsarticle}

\usepackage{xcolor}
\usepackage{graphicx}
\usepackage{lineno,hyperref}
\modulolinenumbers[10]

\journal{Journal of Nuclear Materials and Energy}

\newcommand{\diff}[2]%
{
  \frac{d#1}{d#2}
}
\newcommand{\diffs}[2]%
{
  \frac{d^2#1}{d#2^2}
}
\newcommand{\pdiff}[2]%
{
  \frac{\partial#1}{\partial#2}
}
\newcommand{\pdiffs}[2]%
{
  \frac{\partial^2#1}{\partial#2^2}
}
\newcommand{\pdiffx}[3]%
{
  \frac{\partial^2#1}{\partial#2 \partial#3}
}

\newcommand{\tdiff}[2]%
{
  {d#1}/{d#2}
}
\newcommand{\tdiffs}[2]%
{
  {d^2#1}/{d#2^2}
}
\newcommand{\tpdiff}[2]%
{
  {\partial#1}/{\partial#2}
}
\newcommand{\tpdiffs}[2]%
{
  {\partial^2#1}/{\partial#2^2}
}
\newcommand{\tpdiffx}[3]%
{
  {\partial^2#1}/{\partial#2 \partial#3}
}

\def\beq{\begin{equation}}
\def\eeq{\end{equation}}
\def\beqar{\begin{eqnarray}}
\def\eeqar{\end{eqnarray}}
\def\nn{\nonumber}

\def\gsim{{\lower.3em\hbox{${\>\buildrel > \over \sim\>}$}}}
\def\lsim{{\lower.3em\hbox{${\>\buildrel < \over \sim\>}$}}}

\newcommand{\vect}[1]%
{
  {\bf #1}
}

\def\vpe{V_{||e}}
\def\vpi{V_{||i}}

\def\vE{\vec{V}_{E}}

\def\b0{\vec{b}_{0}}
\def\vb{\vec{b}}

\def\jpar{j_{||}}

\def\grad{\nabla}

\def\dpar{{\partial}_{||}}

\def\d0par{\partial^0_{||}}
\def\hd0par{\hat{\partial}^0_{||}}
\def\gradpar{\grad_{||}}

\def\gradperp{\grad_{\perp}}

\def\1{\perp 1}
\def\2{\perp 2}
\def\p{||}

\def\calA{{\cal A}}
\def\calB{{\cal B}}
\def\krho{{k_{\perp} \rho_{ci}}}

\biboptions{sort&compress}

\begin{document}

\begin{frontmatter}
  \title{Analysis of high-field side plasma instabilities in tokamak edge}
  \author{M.V. Umansky} 
  \address{Lawrence Livermore National Laboratory, Livermore, CA 94550, USA}

  \begin{abstract}
    Balanced double-null configurations are of general interest for
    boundary plasma physics, and they have been proposed for some
    future designs. Experimental observations demonstrate absence of
    plasma fluctuations in tokamak high-field side scrape-off layer in
    a balanced double-null configuration [Smick et al 2013
      Nucl. Fusion 53 023001], and it is commonly assumed that plasma
    instabilities are suppressed on high-field side in the edge plasma
    due to the stabilizing effect of magnetic curvature. At the same
    time, the experimental evidence points to extremely steep plasma
    density profiles on high-field side, which should provide a strong
    instability drive. In the present study, the
    drift-resistive-ballooning mode instability model is investigated
    analytically and numerically to determine the characteristics of
    plasma instabilities, turbulence, and transport in tokamak
    scrape-off layer on high-field side.
  \end{abstract}


\end{frontmatter}


\newpage

\section{Introduction}

Tokamak balanced double-null (DN) configurations are of general
interest for boundary plasma physics, and they have been proposed for
some future designs such as ARIES \cite{Najmabadi1990}, FDF
\cite{Chan2010}, ARC \cite{Kuang2018}, EU-DEMO \cite{AhoMantila2021},
and others. One reason such configurations are attractive is that
using two outer divertors may simplify dealing with the exhaust
power. Another attractive feature is that in balanced DN
configurations the scrape-off-layer (SOL) plasma on high-field side
(HFS) has been found to be quiescent, unlike the low-field side (LFS),
and placing radio-frequency (RF) actuators on HFS may be more
efficient due to expectation of low scattering in the SOL on HFS
\cite{Wallace2015}.

The subject of this study is analysis of plasma linear stability and
turbulence in the SOL on the HFS of a tokamak, in particular for
near-balanced DN magnetic configurations when the HFS SOL does not
directly connect to the LFS SOL.

For a balanced DN configuration, the LFS and HFS of the SOL are
essentially the same in terms of physics. Topologically it is a
slab-like domain with toroidal and poloidal magnetic field; the
difference between the LFS and HFS SOL is basically in the direction
of the magnetic curvature vector with respect to the plasma density
and pressure gradient, which makes the magnetic curvature
destabilizing on LFS and stabilizing on HFS.

There are many experimental observations of LFS SOL on different
tokamaks, indicating ubiquitious presence of turbulent fluctuations
and transport. For the HFS, there are very few measurements. Probably
the best diagnosed HFS SOL is in Alcator C-Mod, where scanning
Langmuir probe measurements were carried out on HFS
\cite{Smick2005,Smick2013,LaBombard2017}. The main observations from
those studies for balanced DN configurations were that (i) HFS SOL is
extremely narrow, and (ii) the level of fluctuations and
fluctuations-induced radial particle flux on HFS is very
low. Theoretical analysis of these observations is the subject of the
present study.



\section{Physics model}

For the model of plasma instabilities in tokamak SOL we use the following plasma fluid equations:

\medskip
\medskip
\medskip

{Density conservation}

\beqar
\pdiff{N_i}{t} = - \vE \cdot \grad N_i
\label{e:phys_density}
\eeqar

{Charge conservation (potential vorticity)}

\beqar
\pdiff{\varpi}{t}  = -  \vE \cdot \grad \varpi
- 2 \omega_{ci} \vb \times \kappa \cdot \grad P_{ei}
- N_i Z_i e \frac{4 \pi V_A^2}{c^2} \gradpar \jpar 
\label{e:phys_vorticity}
\eeqar

{Electron parallel momentum}

\beqar
\frac{e}{m_e} E_{\p} +
\frac{1}{N_i m_e} (T_e \dpar N_i) =
0.51 \nu_{ei} (\vpi - \vpe)
\label{e:phys_elpmom}
\eeqar

where

\beqar
\varpi = N_i e \gradperp^2 \phi 
\nn \\
\vE = c \b0 \times \gradperp \phi /B 
\nn \\
E_{||} = - \dpar \phi
\nn \\ 
%
%
\gradpar F = B \dpar (F/B) \nn
\eeqar

Eqs. (\ref{e:phys_density},\ref{e:phys_vorticity},\ref{e:phys_elpmom}),
with some variations, form the basis of many theoretical studies of
tokamak edge turbulence, pioneered in the early 1990s by Guzdar et
al. \cite{Guzdar1993}. As these equations support drift-resistive and
ballooning instabilities this physics model is often called the
drift-resisitive-ballooning mode (DRBM). Here the DRBM model is
analyzed in a double-periodic slab geometry sketched in
Fig. (\ref{fig_slab_domain}); $x,y,z$ are the field-aligned
coordinates described, e.g., in Ref. (\cite{Umansky2009}).



\section{Linear analysis}

\subsection{Linearized equations}

We assume uniform background temperature $T_{e0}$=const, for the
background plasma density assume constant radial gradient scale length
$\grad N_{i0} = N_{i0} / L_n$, no background ion flow, $\vpi=0$, no
background electric potential, $\phi_0=0$, and weakly varying
background magnetic field B. We assume constant curvature in the
negative radial direction, $\vec{\kappa} = - \hat{e}_x/R$, and $\vb
\times \kappa$ perpendicular to the field within flux surface in the
positive $\grad z$ direction (see the geometry sketch). Then in the
field-aligned $(x,y,z)$ coordinates (both $y$ and $z$ are periodic)
the operator $\vb \times \kappa \cdot \grad$ appears to be $(1/R)
\grad_{z}$. For details of differential operators in the field-aligned
coordinates see, e.g., Ref. (\cite{Umansky2009}).

Then the linearized Fourier-decomposed equations are

\beqar
- i \omega \tilde{N}_i = -\frac{i c k_{\perp}}{B} \frac{N_{i0}}{L_n} \tilde{\phi}
\label{e:phys_density_lin}
\eeqar

\beqar
- i \omega N_{i0} e k_{\perp}^2 \tilde{\phi} =
2 \Omega_{ci} T_{e0} \frac{1}{R_c} i k_{\perp} \tilde{N}_i +
\\ \nn
+ (N_{i0} e)^2 \frac{4 \pi V_A^2}{c^2} i k_{||} \tilde{V}_{||e}
\label{e:phys_vorticity_lin}
\eeqar

\beqar
0 = \frac{e}{m_e} i k_{||} \tilde{\phi} - \frac{T_{e0}}{N_{i0} m_e} i k_{||} \tilde{N}_i
- 0.51 \nu_{ei} \tilde{V}_{||e}
\label{e:phys_elpmom_lin}
\eeqar

\subsection{Linear dispersion relation}

The linearized Fourier-decomposed equations (using ion charge $Z_i$=1)
lead to the dispersion relation

\beq
\omega^2 + \Omega_K^2 + i {\sigma_{||}} (\omega - \omega_{*}) = 0
\eeq

where the standard textbook \cite{Chen1984} notation is used,

\beq
\sigma_{||} 
= 
\left( \frac{k_{||}}{k_{\perp}} \right)^2 
\frac{\Omega_{ci} \omega_{ce}}
{0.51 \nu_{ei}} 
\eeq

\beq
\omega_{*e} = k_{\perp} V_{*e} = k_{\perp} \frac{V_{te}^2}{\omega_{ce} L_n} 
\eeq

\beq
\Omega_K^2 = \pm {\frac{2 C_s^2}{R L_n}}
\eeq

\beq
C_s^2 = {T_{e0}/M_i}
\eeq

\beq
V_{te}^2 = {T_{e0}/m_e}
\eeq

Due to the relative orientation of the magnetic curvature and the SOL
density gradient, on LFS the curvature term is positive,
$\Omega_K^2>0$; and on HFS it is negative, $\Omega_K^2<0$.

\subsection{Solution of linear dispersion relation}

Normalizing each term in the dispersion relation by $\omega_{*}$ we obtain

\beq
\hat{\omega}^2 + \hat{\Omega}_K^2 + i \hat{\sigma}_{||} (\hat{\omega} - 1) = 0
\label{eq_drbm_norm}
\eeq

This is a quadratic equation, and analyzing the coefficients one can
see that for $ \hat{\Omega}_K^2 \ge$0 the equation always has an
unstable solution as it has one root above and one below the real
axis. Indeed,

\beq
\hat{\omega_1} + \hat{\omega_2} = - i \hat{\sigma}_{||}
\eeq

so the real parts of roots are opposite.

Next, using $\hat{\omega_1} = a + b i$, $\hat{\omega_2} = -a + c i$,
where $a,b,c$ are real, we find

\beq
\hat{\omega_1} \hat{\omega_2} = (-a^2 -bc) + (ac-ab) i =
\hat{\Omega}_K^2 - i \hat{\sigma}_{||},
\eeq

so

\beq
bc = -a^2 - \hat{\Omega}_K^2
\eeq

For $\hat{\Omega}_K^2 >$0 (on LFS) the product $bc$ must be negative;
in other words, the imaginary parts of the two roots have opposite
sign, so there is always one stable and one unstable root. However,
for $\hat{\Omega}_K^2 <$0 (on HFS) it can happen that there is no
unstable root, depending on the size of the curvature term. Using
$\hat{\Omega}_K^2 = -1$ in the dispersion relation
Eq. (\ref{eq_drbm_norm}) , one can rewrite it as

\beq
(\hat{\omega}-1)(\hat{\omega}+1+i \hat{\sigma}_{||}) = 0,
\eeq

so one can see that this is a marginal situation when one root is on
the real axis and the other one in the lower half of the complex
plane. For $\hat{\Omega}_K^2 <$-1 both roots are in the lower-half of
the complex plane, so there is no unstable solution anymore. On the
other hand, for $-1 < \hat{\Omega}_K^2 < 0$ there is still an unstable
root, but the growth rate is reduced by the curvature.

\medskip

Next, defining the collisionality parameter

\beq
\calA = \frac{1}{0.51} \sqrt{\frac{M}{m}} (k_{||}^2 \lambda_{ei} L_n),
\eeq

the curvature parameter

\beq
\calB = \sqrt{\frac{2 L_n}{R}},
\eeq

and the normalized nondimensional wavenumber

\beq
\xi_{\perp} = k_{\perp} \rho_{ci},
\eeq

one can write the dispersion relation Eq. (\ref{eq_drbm_norm}) as

\beq
\hat{\omega}^2 +
i \frac{\calA}{\xi_{\perp}^3} \hat{\omega} +
\frac{\calB^2}{\xi_{\perp}^2} - i \frac{\calA}{\xi_{\perp}^3} = 0
\label{eq_drbm_krho}
\eeq

The solutions of Eq. (\ref{eq_drbm_krho}) are shown in
Fig. (\ref{fig_drbm_krho}) where the instability growth rate
$Im(\hat{\omega})$ is plotted against $\xi_{\perp}$ for several values of
parameters $\calA,\calB$. One can see in the
Fig. (\ref{fig_drbm_krho}) that depending on the parameters
$\calA,\calB$ there is a window of instability. For negative
${\calB}^2$ the growth rate is reduced compared to the no-curvature
case $\calB=0$, and there is a cut-off value $\xi_{\perp} = |\calB|$
below which the instability is suppressed.



\section{Numerical simulations}

Numerical solutions of
Eqs. (\ref{e:phys_density},\ref{e:phys_vorticity},\ref{e:phys_elpmom})
is carried out with SOLT3D \cite{Umansky2024} which is a plasma
simulation code implemented in the BOUT++ framework
\cite{Dudson2009}. Similar to the linear analysis, the numerical
calculations are performed in the slab geometry using field-aligned
coordinates sketched in Fig. (\ref{fig_slab_domain}), where both $y$
and $z$ coordinates are periodic.

First, we verify the linear dispersion relation in the code. Figure
(\ref{fig_bmrk_krho}) shows the solution to Eq. (\ref{eq_drbm_krho})
and the results from SOLT3D, which appear in full consistency, for a
particular choice of parameters $\calA$=468 and $\calB^2$=-8.9 where
the local theory is valid. Note that for C-Mod DN HFS, the relevant
values of the nondimensional parameters would be $\calA \approx 0.1$
and $\calB^2 \approx -0.003$, using $L_n=$0.1 cm, $R=$70 cm,
$\lambda_{ei}=$10 cm, $k_{||}=$0.03 cm$^{-1}$, estimated from
experimental data in Ref. \cite{Smick2005}.

Next, nonlinear simulations are carried out where the system relaxes
to saturated turbulence. This is still done for shallow radial plasma
profles, far from the experimental situation where the plasma density
drops rapidly across the separatrix resulting in an extremely thin
scrape-off layer on the HFS with very sharp radial plasma profiles in
balanced double-null \cite{LaBombard2017,Smick2005}. However, with the
chosen parameters, the system is much more tractable for both analytic
treatment and the simulations, which still allows capturing some
important features of tokamak plasma on HFS; extending the analysis to
realistic HFS plasma profiles is a planned extension of the present
study.

For the present numerical simulations, the parameters are: $R$=1 m,
$T_{e0}$=100 eV, $N_{i,sep}$=0.5$\times$10$^{20}$ m$^{-3}$, $B_t$=1 T,
$B_p$=0.1 T. Three cases are considered, with $L_n$=2.25 m, 1.125 m,
and 0.5625 m.

In the numerical simulations, the average plasma density profile is
maintained fixed while the electric potential and vorticity profiles
evolve freely. The system evolving in time goes through the linear
growth phase and settles in a steady turbulence state.

One quantity of primary interest in the simulations is the radial
density flux $\Gamma_x$ in the steady turbulence state. We calculate
$\Gamma_x$ for three different values of the plasma density scale
length $L_n$. It is found that the fluctuations amplitude and the
time-average flux $\left<\Gamma_x\right>$ grow strongly with $1/L_n$,
see Fig. (\ref{fig_flux_scaling}) where the symbols show the results
from the simulations, the dashed line shows the power-law fit $\Gamma
= 3 \times 10^{14}/L_n^4$ and the dot-dashed line shows the
exponential fit $\Gamma = 4 \times 10^{12} \exp(4/L_n)$ (in MKS
units).

Another interesting observation is that the probability distribution
function (PDF) of $\Gamma_x$ undergoes significant changes as $1/L_n$
grows, going from nearly symmetric distrubution to a strongly skewed
one, see Fig. (\ref{fig_flux_skewness})



\section{Discussion}

The strong dependence of the turbulent radial flux $\Gamma_x$ on the
plasma density radial scale length $L_n$ suggests that the radial
plasma profile is governed by marginal stability physics, so the
resulting level of fluctuations may be small.

Indeed, for the plasma profile at the separartix, one can make an
estimate from the condition $\grad \cdot \vec{\Gamma} \approx$ 0,

\beq
\frac{\Gamma_x}{L_n} \sim \frac{C_s N_i}{L_{||}},
\eeq

so for an equilibrium one needs

\beq
{\Gamma_x} = \alpha L_n,
\eeq

where $\alpha=C_s N_i/L_{||}$ is a numerical factor.

On the other hand, the numerical results in
Fig. (\ref{fig_flux_scaling}) show rapid growth of $\Gamma_n$ with
$1/L_n$, which can be approximated, e.g., by an exponential scaling,

\beq
{\Gamma_x} = \beta \exp(\lambda /L_n),
\eeq

where $\beta$, $\lambda$ are numerical factors.

Thus, the equilibrium density scale length $L_n$ is defined by the
solution to the equation

\beq
\alpha L_n = \beta \exp(\lambda /L_n),
\label{eqn_marginal_stability}
\eeq

which has a single root, and the resulting $L_n$ (and the
corresponding level of turbulent fluctuations) will depend on the
numerical values of parameters $\alpha$, $\beta$, and $\lambda$. Of
course, based on the present simulations with $L_n \sim$ 1 m it would
be meaningless to extrapolate to realistic $L_n \sim$ 1 mm.

%
%

\medskip

The changes of $\Gamma_x$ PDF with $L_n$ shown in
Fig. (\ref{fig_flux_skewness}) is an interesting feature suggesting
coupling of unstable and damped modes in the steady state
turbulence. It is known that stable (damped) modes may play a
significant role in plasma turbulence \cite{Fraser2018}. One can make
an argument in the spirit of the quasilinear plasma theory, noticing
that unstable modes correspond to radially outward radial plasma
density flux while the damped modes correspond to radially inward
density flux.

Indeed, consider the linearized density evolution equation,
Eq. (\ref{e:phys_density_lin}), in the Fourier decomposed form,

\beq
-i \omega \tilde{N}_i = \tilde{V}_E N_{i0}/L_n,
\eeq

\medskip

Then,

\beq
\tilde{V}_E = - i \omega (L_n/N_{i0}) \tilde{N}_i,
\eeq

and assuming that the mode beats with itself, for the average density
flux we find,

\beq
\left<\Gamma_x \right> =
\frac{1}{2} Re(\tilde{N}_i \tilde{V}_{E}^{*}) =
\frac{1}{2} |\tilde{N}_i|^2 (L_n/N_{i0}) \gamma
\eeq

where $\gamma=Im(\omega)$ is the mode growth rate.

Thus, unstable modes give rise to outward radial density flux and
damped modes give rise to inward flux, which may be relevant to the
$\Gamma_x$ PDF in Fig. (\ref{fig_flux_skewness}). Indeed, as can be
seen from the linear analysis, the instability domain corresponds to
$k_{\perp} \rho_{ci} > \sqrt{2 L_n /R}$; so, for large $L_n$ there is
a large pool of damped modes. Consistent with that, as
Fig. (\ref{fig_flux_skewness}) shows, for large $L_n$ there is a
significant contribution of negative fluctuations of turbulent flux.

It is interesting to note that for the parameters used in the present
numerical simulations, the turbulent flux on LFS is orders of
magnitude higher than that on HFS. This is not too surprising, given
that on LFS the stabilizing curvature term becomes destabilizing and
there is robust instability for all wavenumbers.



\section{Conclusions}

Linear and nonlinear analysis is conducted using the DRBM instability
model for tokamak SOL plasma on HFS in a balanced DN configuration
where the magnetic curvature has stabilizing effect. Experimental
evidence on HFS points to abscence of fluctuations and to steep radial
gradients of plasma density and pressure on the separatrix. On the
other hand, the present analysis indicates that steep gradients should
favor the instability drive over the magnetic curvature stabilization,
and some level of fluctuations must be present on HFS, although the
amplitude may be small due to marginal stability type constraints.
For quantitative comparison with the experiment, apparently a more
complete model will be needed, using realistic radial plasma
profiles. Furthermore, the extended model should include mechanisms
that may suppress the instability, such as the finite Larmor radius
(FLR) effects, the strongly varying background radial electric field,
and the X-point geometry with magnetic field shearing. In view of the
ideas to place RF actuators on the tokamak HFS \cite{Wallace2015}, an
important question for further investigation is whether edge plasma
fluctuations can be destabilized on HFS by the RF antenna field.


\section*{Acknowledgements}

The author gratefully acknowledges discussions with B.~I. ~Cohen and
J. R. Myra.  This work was performed under the auspices of the US
Department of Energy by Lawrence Livermore National Security, LLC,
Lawrence Livermore National Laboratory, under Contract
DE-AC52-07NA27344.



\clearpage
\newpage
\medskip
\medskip
\medskip
\bibliography{hfside_refs}

\clearpage
\clearpage

\section{Figure captions}

\begin{figure}[h] 
\caption{
The slab domain representing SOL. The coordinates $x,y,z$ shown are the field-aligned coordinates used in the calculation.
\hfill
}
\label{fig_slab_domain}
\end{figure}

\begin{figure}[h] 
\caption{
Normalized DRBM growth rate $\hat{\omega}$ vs. the nondimensional
parameter $\xi_{\perp} = \krho$ for several different values of
parameters $\calA, \calB$ defined in the main text.
\hfill
}
\label{fig_drbm_krho}
\end{figure}

\begin{figure}[h] 
\caption{
Data points from linear simulations shown with corresponding analytic
results for the wavenumber scan.
\hfill
}
\label{fig_bmrk_krho}
\end{figure}

\begin{figure}[h] 
\caption{
Scaling of turbulent radial density flux with the density scale
length. The symbols show results from the simulations, the dashed line
shows the power-law fit and dot-dashed line shows the exponential fit
explained in the main text.
\hfill
}
\label{fig_flux_scaling}
\end{figure}

\begin{figure}[h] 
\caption{
The probability density function (PDF) of the radial density flux for
different values of the density scale length $L_n$.
\hfill
}
\label{fig_flux_skewness}
\end{figure}

\clearpage



{
\LARGE{Fig. 1}
\begin{figure}
\includegraphics[scale=0.5]{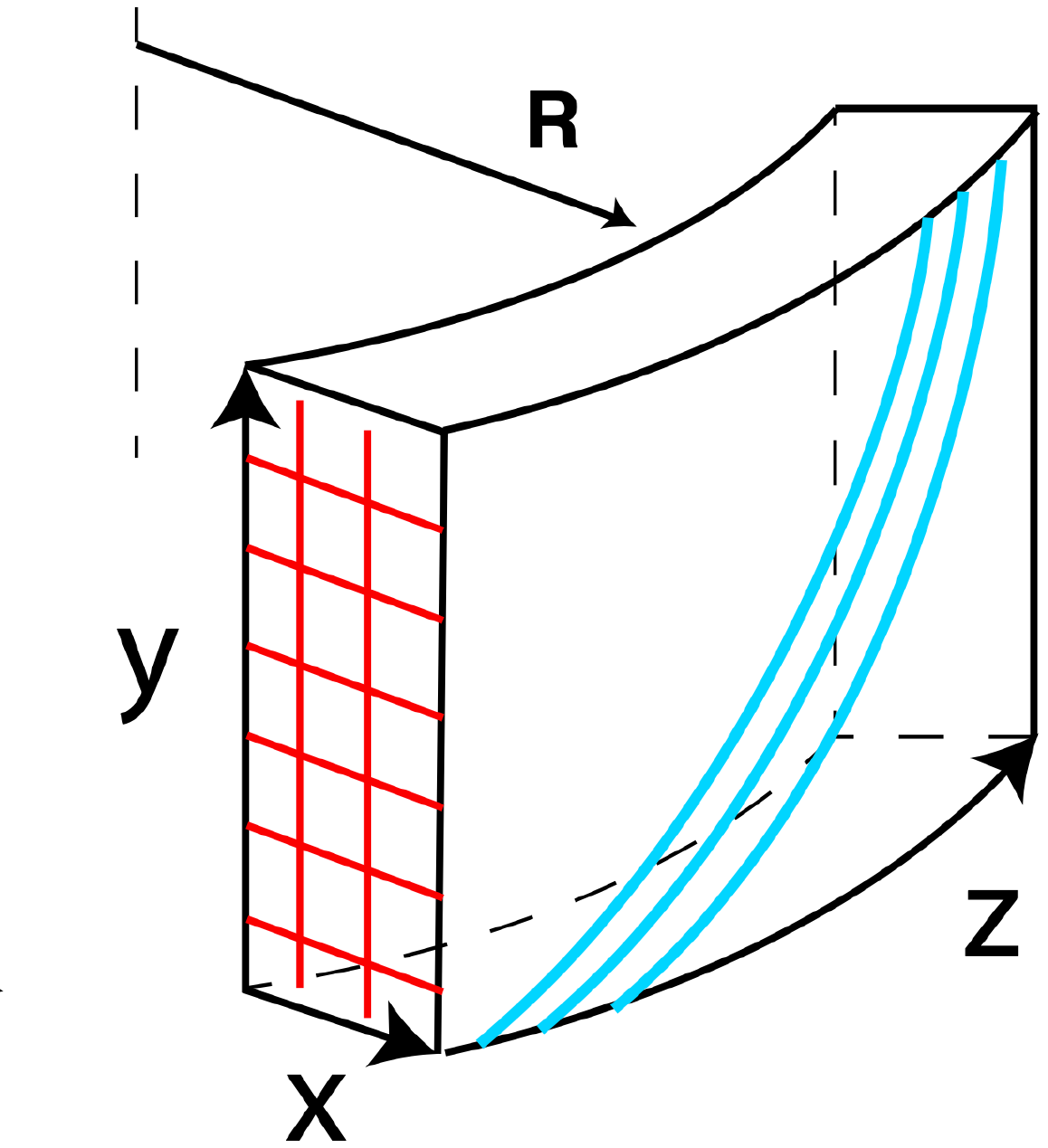}
\end{figure}
\clearpage
}

{
\LARGE{Fig. 2}
\begin{figure}
\includegraphics[scale=1.0]{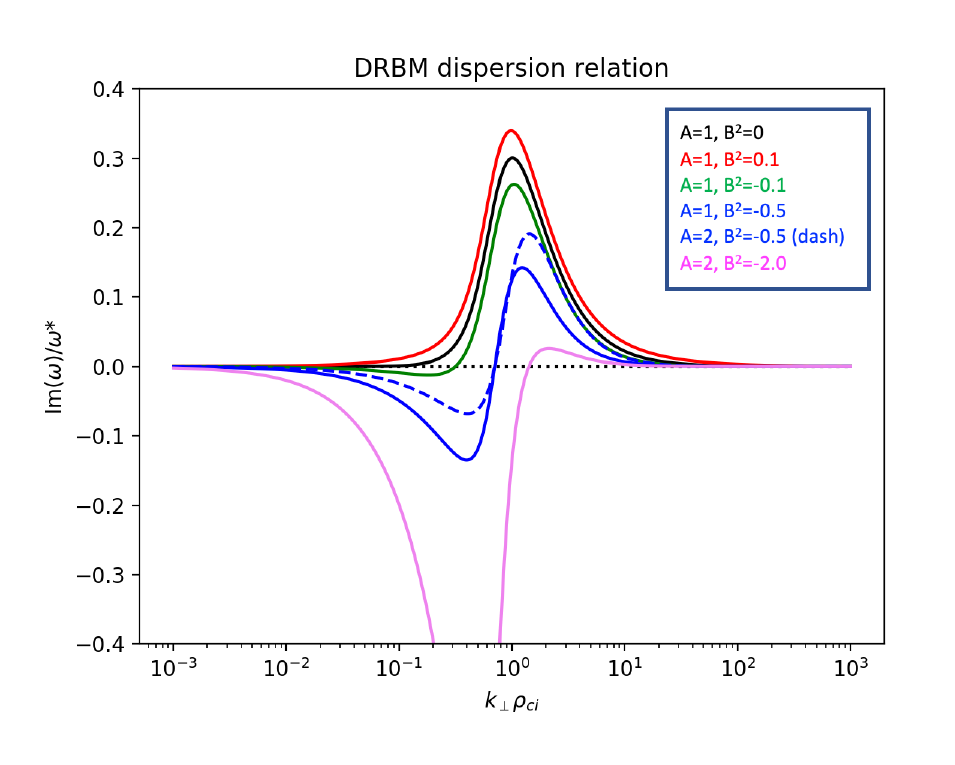}
\end{figure}
\clearpage
}

{
\LARGE{Fig. 3}
\begin{figure}
\includegraphics[scale=0.5, angle=-90]{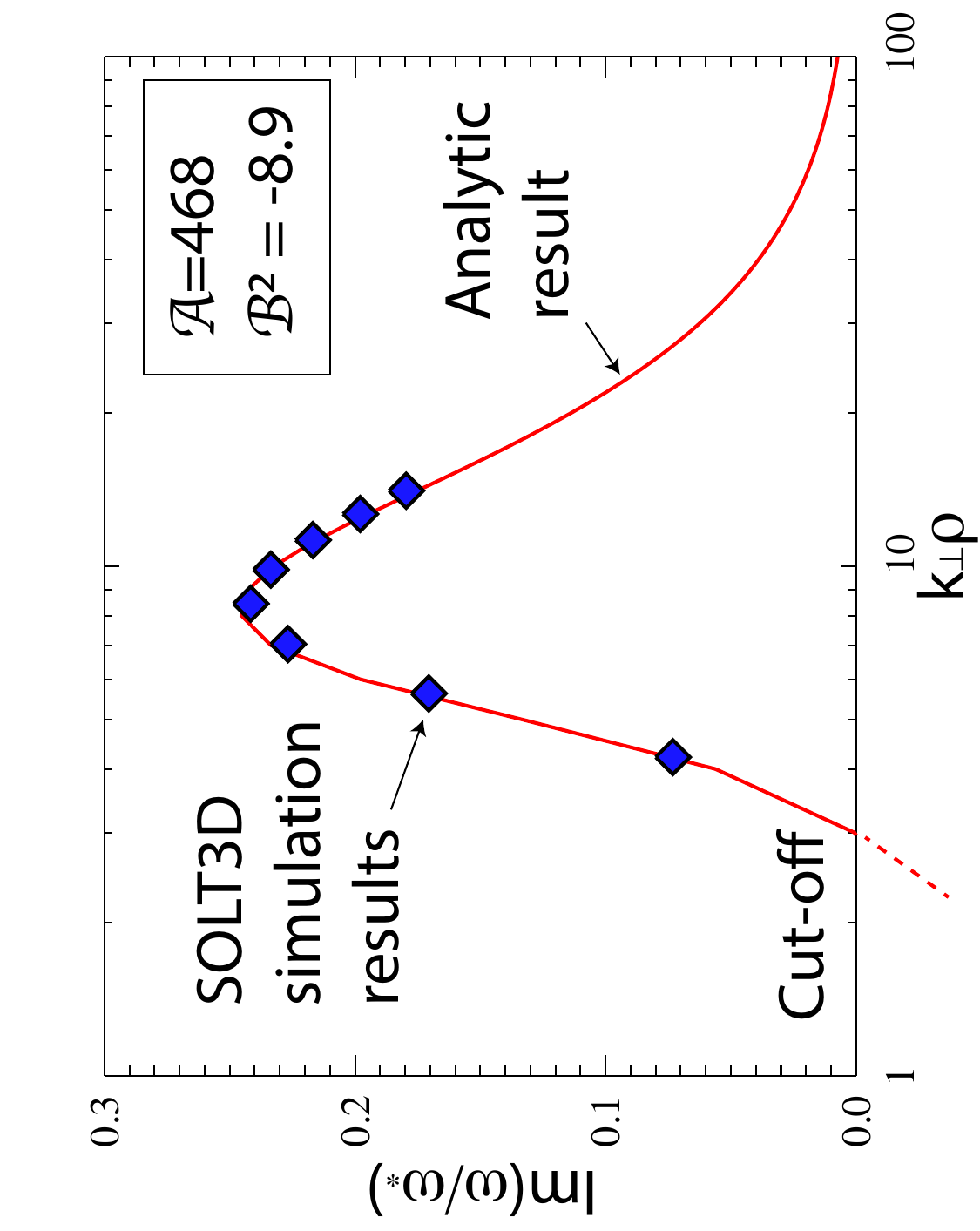}
\end{figure}
\clearpage
}

{
\LARGE{Fig. 4}
\begin{figure}
\includegraphics[scale=0.75]{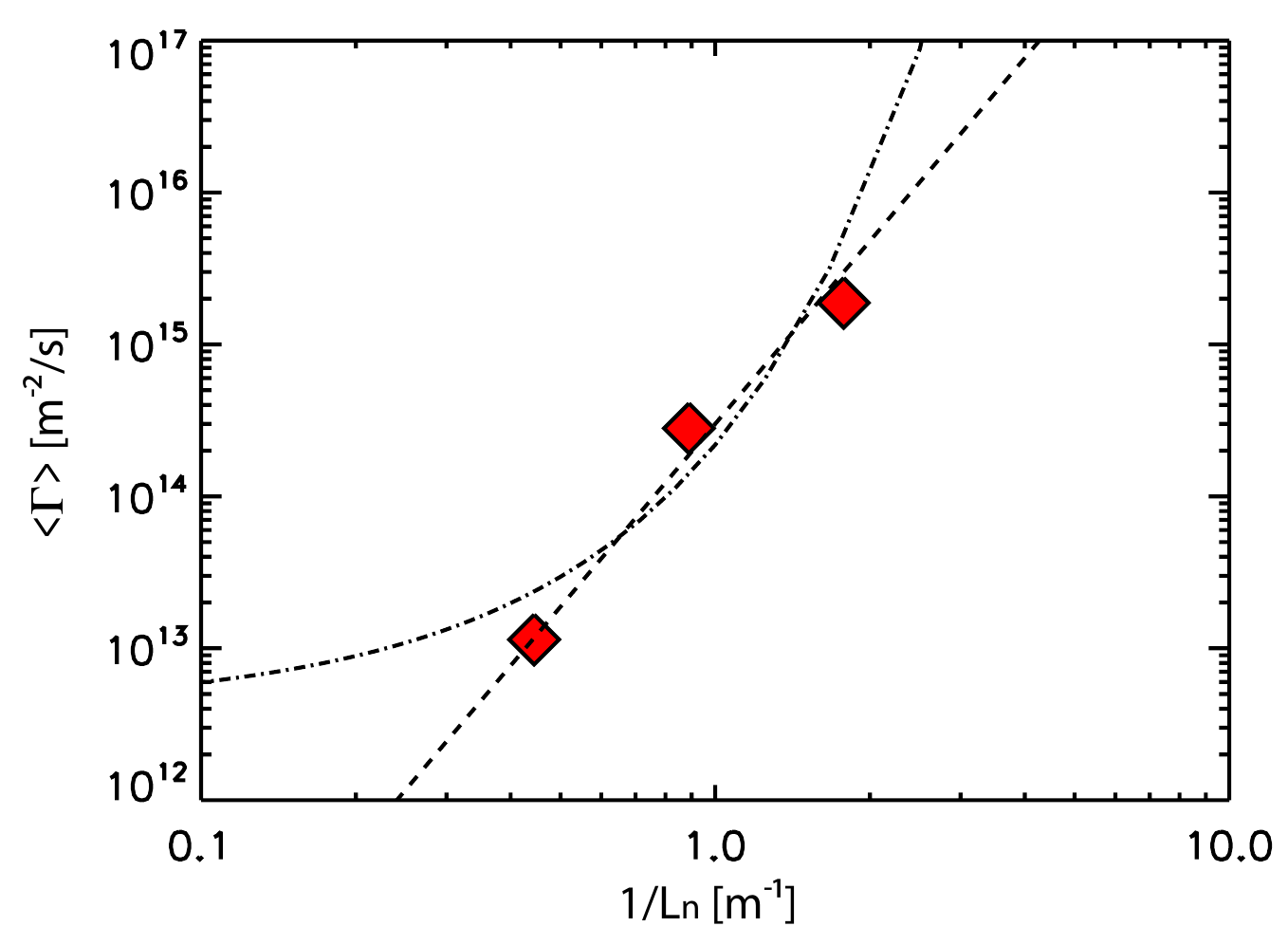}
\end{figure}
\clearpage
}

{
\LARGE{Fig. 5}
\begin{figure}
\includegraphics[scale=0.40]{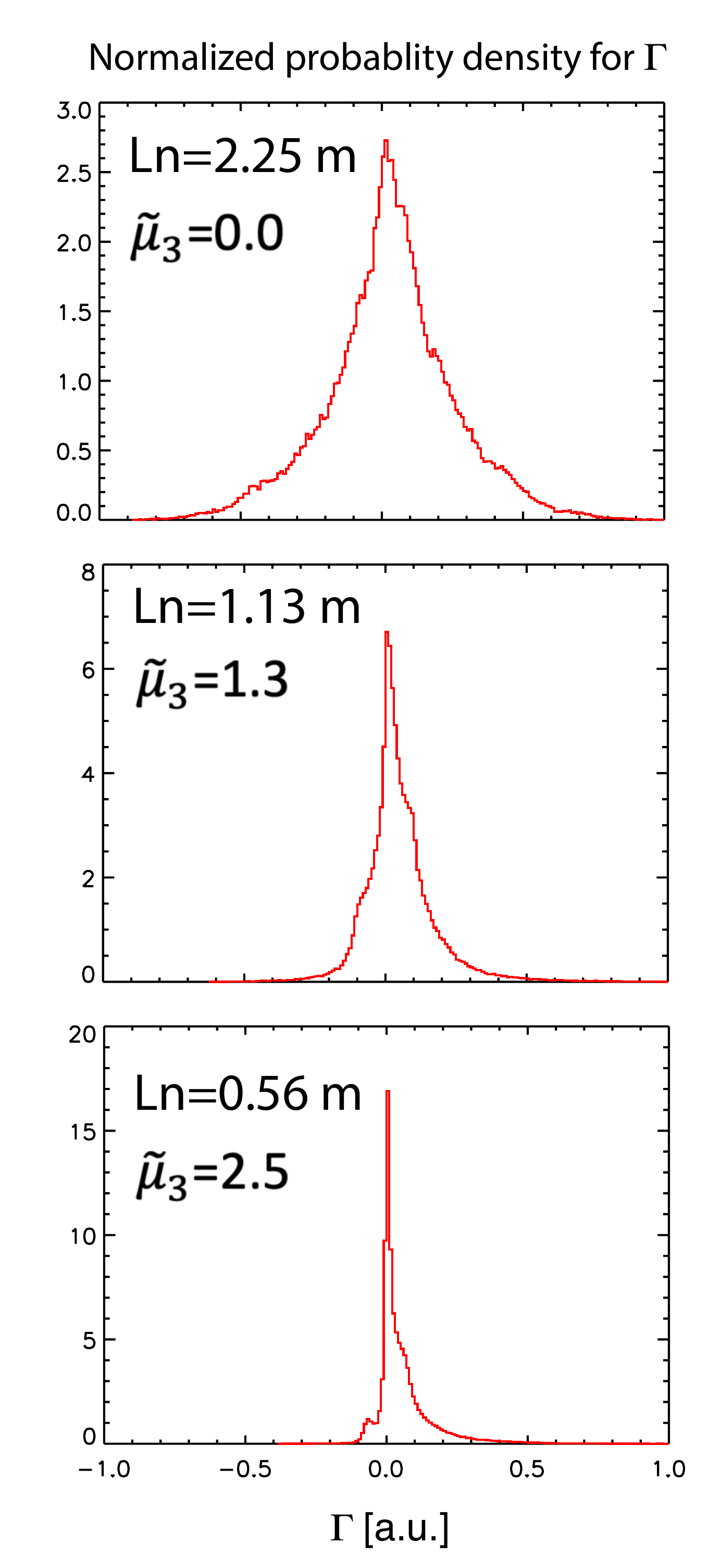}
\end{figure}
\clearpage
}


%
%
\end{document}